\documentclass[aip,jcp,graphicx,amsmath,amssymb,reprint]{revtex4-1}
\usepackage{graphicx}
\usepackage{dcolumn}
\usepackage{bm}
\usepackage{color}
\draft 

\begin{document}


\title{Molecular Dynamics Simulation of Soundwave Propagation in a Simple Fluid} 



\author{Yuta Asano}\email{yuta.asano@issp.u-tokyo.ac.jp}
\affiliation{Institute for Solid State Physics, The University of Tokyo, Kashiwa, Chiba 277-8581, Japan}
\author{Hiroshi Watanabe}%
\affiliation{Department of Applied Physics and Physico-Informatics, Keio University, Yokohama 223-8522, Japan}
\author{Hiroshi Noguchi}
\affiliation{Institute for Solid State Physics, The University of Tokyo, Kashiwa, Chiba 277-8581, Japan}



\date{\today}

\begin{abstract}
 A molecular dynamics (MD) simulation was performed to study the propagation of soundwaves in a fluid. Soundwaves are generated by a sinusoidally oscillating wall and annihilated by a locally applied Langevin thermostat near the opposite wall. The waveform changes from sinusoidal to sawtooth with increasing wave amplitude. For low-frequency sounds, the simulation results show a very good agreement with Burgers's equation without any fitting parameters. In contrast, for high-frequency sounds, significant deviations are obtained because of acoustic streaming. The speed of sound can be directly determined from the Fourier transform of a waveform with high accuracy. Although obtaining the attenuation rate directly from the simulation results is difficult because of the nonlinear effects of the wave amplitude, it can be estimated via Burgers's equation. The results demonstrate that MD simulations are a useful tool for the quantitative analysis of soundwaves.
\end{abstract}


\maketitle 

\section{\label{sec:introduction}Introduction}
Soundwaves are a familiar phenomenon in our daily lives. They include the sound of voices and musical instruments, and they propagate in various media: solids, liquids, and gases. Thus, soundwaves have a wide range of applications in not only engineering but also medical devices and food processing. Soundwaves are usually generated by an object vibrating in a medium.
However, they can be generated in the absence of vibrating objects, such as wind noise from powerlines; this is known as the Aeolian sound~\cite{phillips56}. When soundwaves are propagating in a fluid medium, they are called fluid sounds. In vehicles, such as automobiles and aircraft, and energy delivery systems, such as turbines and pipelines, fluid sound is a significant problem because it causes noise and vibration. Particularly for fluid machinery, cavitation can cause erosion owing to the shock waves produced upon bubble collapse~\cite{sap17}. In many industrial flows, molecular-scale dynamics such as cavitation have a significant effect on macroscopic properties. Therefore, analyzing sound propagation at the molecular scale is of great importance for engineering applications.

A fundamental problem for soundwave propagation is soundwave generation from a sinusoidally oscillating flat plate. The shockwave formation process and generation of acoustic streaming, including nonlinearity, have been analyzed with the Navier--Stokes equation~\cite{blackstock64, blackstock66, iy94, iy97}. Moreover, this setup has been used to study molecular motion in microelectromechanical systems~\cite{ht98,mdm19}. Because the characteristic length of the system is comparable with the length of the molecular mean free path, the propagation characteristics of a soundwave in rarefied gases cannot be described by the Navier--Stokes equation.
Therefore, mesoscale approaches such as the Boltzmann equation and direct simulation Monte Carlo method have been used to analyze sound propagation characteristics~\cite{ta13, sgc98, hg01}. 
Recent improvements in computational power have led to the application of molecular dynamics (MD) simulations to studying sound propagation in rarefied gases~\cite{yano12, azh18a, azh18b, bch19}. These studies focused on the attenuation coefficients and waveform distortions near the oscillating plate, mainly for linear waves. Shockwave propagation in rarefied gases was also simulated~\cite{yano12}.
On the other hand, for solids and dense fluids, MD simulations of shock waves have been reported~\cite{hhm80, holian88,zzn99}. The shock wave is generated by compression such as shrinkage of the computational cell and a piston potential. Recently, the effects of shock waves on protein fibrils have been simulated using MD~\cite{oi14, vdn16}.

However, there have been virtually no MD simulations in the literature of soundwave propagation in a fluid, except for rarefied gases and shock waves as mentioned above, because the Navier--Stokes equation allows for a sufficiently accurate analysis of soundwaves in a Newtonian fluid. However, complex fluid behaviors as seen in industrial flows, such as polymer solutions and cavitation, are difficult to address with a continuum model. The quantitative analysis of soundwaves at the molecular scale should help elucidate various phenomena related to complex fluids. Moreover, an MD simulation provides direct analysis of the interaction between a fluid and a structure, such as the heat exchange between a solid wall and the fluid; this is vital for the mechanical design. In this study, we performed an MD simulation of soundwaves in a simple fluid and validated the results by comparison with hydrodynamic calculations. We adopted Burgers's equation as the fluid model for comparison with the MD simulation results. Because all parameters required by Burgers's equation were determined through different MD simulations, the comparison did not require any fitting parameters. The two sets of results should agree if the continuum description is valid. 
The rest of this paper is organized as follows. Section II describes the simulation model and method. Section III compares the waveforms of the MD and hydrodynamic simulations. Section IV presents a summary and discussion of the results.
\section{\label{sec:Method}Method}
\subsection{\label{sec:MethodMD}MD simulation}
We adopted the smoothed-cutoff Lennard-Jones (LJ) potential for the interactions between fluid particles. The potential function $\phi$ is
\begin{eqnarray}
  \phi(r) &=& \left\{ \begin{array}{ll}
    \phi_{\rm LJ}(r) - \phi_{\rm LJ}(r_{\rm c}) - (r-r_{\rm c})\phi'_{\rm LJ}(r_{\rm c}) & (r\le r_{\rm c}) \label{eq:ljsf} \\
    0                                                                              & (r>r_{\rm c})
  \end{array} \right. ,\\
  \phi_{\rm LJ} &=& 4\epsilon\left[ \left(\frac{\sigma}{r}\right)^{12} - \left(\frac{\sigma}{r}\right)^6 \right], 
\end{eqnarray}
where $r$ is the inter-particle distance and $\epsilon$ and $\sigma$ are the energy and length scales, respectively. The prime in Eq.~(\ref{eq:ljsf}) represents the derivative with respect to $r$, and $r_{\rm c}=2.5\sigma$ is the cutoff distance of the potential function.
In this paper, Eq.~(\ref{eq:ljsf}) represents the potential of an LJ particle. All physical quantities are in LJ units, that is, the observables are expressed in units of energy $\epsilon$, length $\sigma$, and time $\tau=\sigma\sqrt{m/\epsilon}$, where $m$ is the mass of the LJ particles.
\begin{figure}[h]
  \includegraphics{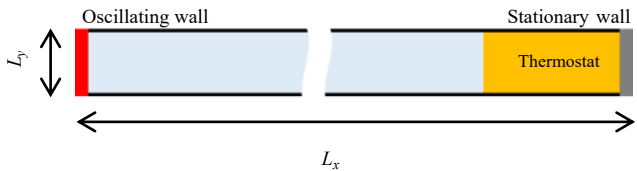}
  \caption{\label{fig:box} Schematic view of the computational domain. The oscillating and stationary walls are indicated by the red and gray regions, respectively. The Langevin thermostat is applied in the yellow region ($39~000 < x < 40~000$).}
\end{figure}

The simulation box is a rectangular parallelepiped with the dimensions $L_x\times L_y\times L_z=40000\times 12.5\times 12.5$, as shown in Fig.~\ref{fig:box}. The periodic boundary condition is adopted for the $y$ and $z$ directions. The fluid is enclosed by two walls in the $x$ direction; the left wall was the sound source, and the right wall was fixed in a position. Both walls were modeled by fixing LJ particles on a face-centered orthorhombic lattice with a density of $0.79$. The left wall vibrates in the $x$ direction as an oscillating plate.
The frequency and amplitude ranges of the oscillation are $0.001 \le f \le 0.004$ and $2.5\le A \le 20$, respectively.
The position of particles on the wall is given by
\begin{eqnarray}
  x_{{\rm w}i}(t)&=&A\sin\left(2\pi f t\right)+x_{{\rm w}i}(0),
\end{eqnarray}
where $x_{{\rm w}i}(t)$ is the position of the $i$th particle of the oscillating wall.
To annihilate the traveling wave and equilibrate the fluid, a Langevin thermostat is applied to the part of the simulation box, shown in yellow in Fig.~\ref{fig:box}. 
The friction coefficient of the thermostat is increased linearly from $0.0001$ to $0.1$ in the region of $39~000<x<39~500$ and is maintained at 0.1 for the rest of the yellow region.
We previously used a similar Langevin thermostat annihilation to remove the flow history of the K\'arm\'an vortex and cavitation~\cite{awn18,awn20}.

In the initial state, the fluid particles are randomly located with zero overlaps within the simulation box. The initial velocities of the fluid particles are given randomly according to the Maxwell velocity distribution. The temperature $T$ is fixed to $T = 2$, where the Boltzmann constant $k_{\rm B}$ is omitted.
A low fluid density of $\rho = 0.1$ is used with $623~338$ particles in total.
We use LAMMPS (Large-scale Atomic/Molecular Massively Parallel simulator)~\cite{plimpton95} to perform numerical integration for up to $30~000~000$ steps with a time step of $0.004$.
We adopt the velocity Verlet algorithm for the time integration. The error bars are estimated from three or more independent runs.

To obtain the waveforms, the simulation box is divided into small cells with the dimensions of $10 \times 12.5 \times 12.5$, and the velocities of particles in each cell are averaged. The time-series data are classified by the phase of the sound source, and the average value for each phase is calculated. Each phase is divided into $20$ sections [time period: $1/(20f)$].
Similarly, the local densities are averaged over time and independent samples for each phase of the sound source.

\subsection{\label{sec:Burgers}Burgers's equation}
We adopted Burgers's equation~\cite{burgers48} to describe the hydrodynamic behavior of LJ particles.
Burgers's equation approximates the Navier--Stokes equation up to the second order of the acoustic quantities and assumes that the soundwave is a one-dimensional traveling wave,
\begin{eqnarray}
  \frac{{\rm \partial}p_{\rm a}}{{\rm \partial}x}
  + \frac{1}{c_0}                      \frac{{\rm \partial}p_{\rm a}}{{\rm \partial}t}
  - \frac{b}{2c_0}                     \frac{{\rm \partial^2}p_{\rm a}}{{\rm \partial}t^2}
  - \frac{\beta p_{\rm a}}{\rho_0 c_0^3} \frac{{\rm \partial}p_{\rm a}}{{\rm \partial}t}
  =0, \label{eq:burgers}
\end{eqnarray}
where $p_{\rm a}$ is the fluctuation component of the pressure and $\rho_0=0.1$ is the density of the stationary fluid.
$c_0$, $b$, and $\beta$ are the speed of sound, attenuation parameter, and nonlinear parameter, respectively.
Although Burgers's equation cannot describe strongly nonlinear phenomena such as the acoustic flow, it can describe soundwaves with a relatively small nonlinearity.
The strength of the nonlinearity is measured in terms of the acoustic Mach number $Ma=2\pi f A/c_0$, which has a range of $0.008<Ma<0.26$ for the amplitude and frequency used in this study.
Within the limit of small nonlinearity, the amplitude of soundwaves decays exponentially according to $\exp\left(-\alpha_0 x\right)$; this is because of dissipation by viscosity and heat transfer in classical theory~\cite{ll89}.
The attenuation parameter $b$ is related to the attenuation coefficient $\alpha_0$ in classical theory by
\begin{eqnarray}
  \alpha_0 = \frac{b\omega^2}{2c_0}, \label{eq:alpha}
\end{eqnarray}
where $\omega=2\pi f$ is the angular frequency. This is obtained by substituting $p_{\rm a}(x,t) = p_{{\rm a}0}\exp[-\alpha_0x+ {\rm i}\omega (t - x/c_0)]$ into Eq.~(\ref{eq:burgers}) while neglecting the last term because $p_{\rm a} \ll 1$.
The parameters $c_0$, $b$, and $\beta$ are obtained as follows:
\begin{eqnarray}
  c_0&=&\sqrt{\frac{K_S}{\rho_0}},\label{eq:c0}\\
  b&=&\frac{1}{\rho_0c_0^2}\left[\left(\zeta+\frac{4}{3}\eta \right) + \kappa\left(\frac{1}{c_V} - \frac{1}{c_p}\right)\right],\label{eq:b}\\
  \beta&=&1+\frac{B_2}{2B_1}.
\end{eqnarray}
$K_S$ is the adiabatic bulk modulus. $\zeta$ and $\eta$ are the bulk viscosity and shear viscosity, respectively. $\kappa$, $c_V$, and $c_p$ are the thermal conductivity, isochoric specific heat, and isobaric specific heat, respectively. $B_1$ and $B_2$ are the first and second adiabatic differential coefficients of the pressure $p$ with respect to the density $\rho$,
\begin{eqnarray}
  B_1&=&\rho_0\left(\frac{{\rm \partial}p}{{\rm \partial}\rho}\right)_S, \label{eq:A}\\
  B_2&=&\rho_0^2\left(\frac{{\rm \partial}^2p}{{\rm \partial}\rho^2}\right)_S, \label{eq:B}
\end{eqnarray}
where $S$ is the entropy.
These quantities are estimated through MD simulations, as described in the Appendix, 
and the parameters in Eq.~(\ref{eq:burgers}) are determined to be $c_0=1.95$, $b=1.66$, and $\beta=1.58$.

Burgers's equation is numerically integrated by using the central difference for the time direction and the second-order backward difference for the spatial direction~\cite{mohamed19}. The boundary conditions are set as follows:
\begin{eqnarray}
  p_{\rm a}(x=0, t)&=& p_{{\rm a}0}\sin \left(2\pi f t\right),\\
  p_{\rm a}(x, t=0)&=&0,\\
  p_{\rm a}(x, t=t_{\rm max})&=&0.
\end{eqnarray}
where $t_{\rm max}=250~000$ and the widths of the spatial and temporal  discretizations are $\Delta x=10$ and $\Delta t=6.25$, respectively.
The amplitude $p_a$ of the sound source is set to the same values as those of the MD simulations according to the plane wave relation $p_{{\rm a}0}=2\pi f A \rho_0 c_0$.

\section{\label{sec:RESULTS}Results}
\begin{figure}[h]
  \includegraphics{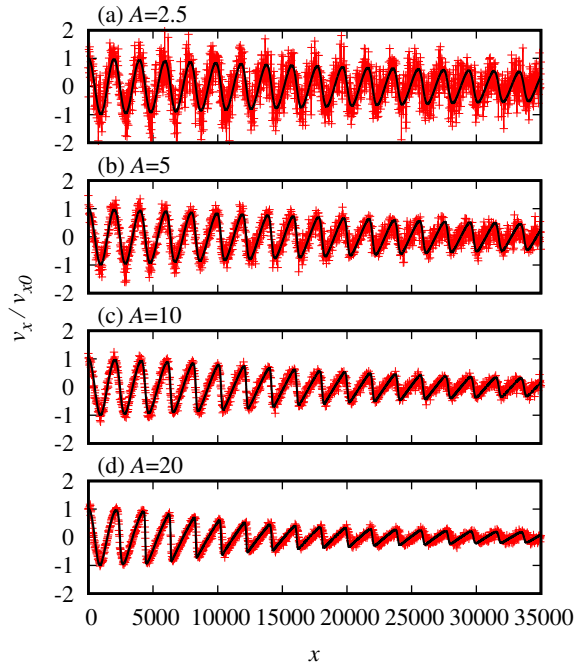}
  \caption{\label{fig:wave_amp}
    Waveform of the LJ fluid at $\rho=0.1$ with plate oscillation amplitudes of (a) $A=2.5$, (b) $A=5$, (c) $A=10$, and (d) $A=20$. The frequency of the plate oscillation is $f=0.001$. The red lines with ($+$) and the black lines represent the MD simulation results and the numerical solution of Burgers's equation, respectively.}
\end{figure}

Figure~\ref{fig:wave_amp} shows the dependence of the waveform on the amplitude at the low frequency $f = 0.001$.
At a small amplitude, the soundwaves propagate with a sinusoidal waveform; however, they deviate to a sawtooth waveform as the amplitude increases. The sawtooth waveform starts to appear closer to the sound source as the amplitude increases.
The black lines in this figure represent the numerical solution of Burgers's equation.
At $f=0.001$, the two waveforms show a very good agreement at the amplitudes considered in this study.

\begin{figure}[h]
  \includegraphics{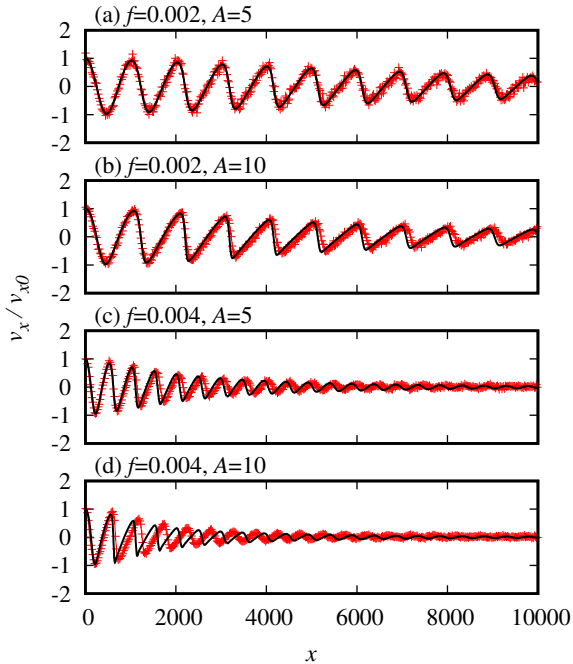}
  \caption{\label{fig:wave_freq}
    Waveform of the LJ fluid at $\rho = 0.1$: (a) frequency $f=0.002$ and amplitude $A=5$, (b) $f=0.002$ and $A=10$, (c) $f=0.004$ and $A=5$, and (d) $f=0.004$ and $A=10$. The red lines with ($+$) and the black lines represent the MD simulation results and the numerical solution of Burgers's equation, respectively.} 
\end{figure}

Figure~\ref{fig:wave_freq} shows the dependence of the waveform on the frequency.
When the frequency is increased to $f=0.002$, the waveforms of the MD simulation and Burgers's equation are mostly consistent; slight deviations are observed far from the sound source when the amplitude is large at $A=10$.
At the much higher frequency of $f=0.004$, the discrepancy starts to appear near the sound source.
The discrepancy increases with the amplitude.

\begin{figure}[h]
  \includegraphics{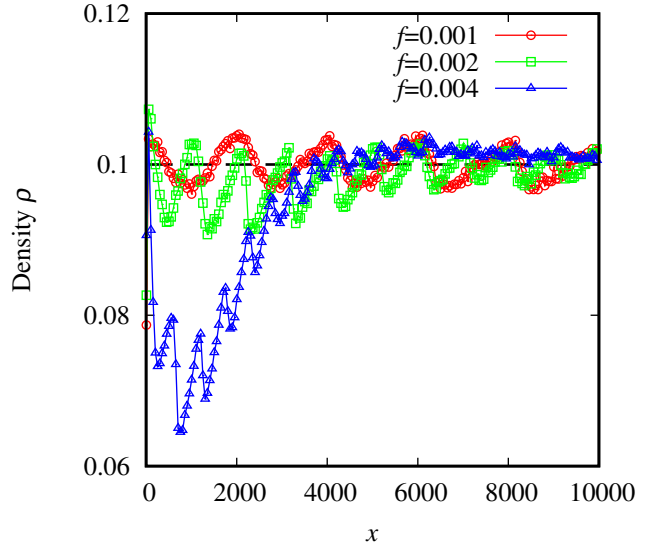}
  \caption{\label{fig:density} Density profile of the LJ fluid at $\rho=0.1$ and  $A=10$ for (a) $f=0.001$, (b) $f=0.002$, and (c) $f=0.004$.}
\end{figure}

Figure~\ref{fig:density} shows the density profiles from the MD simulation. 
At a low frequency, the local density oscillates uniformly around the average density. 
However, the density becomes non-uniform as the frequency is increased, and the density decreases significantly near the source. 
Because Burgers's equation does not consider the spatial dependence of the mean density, this non-uniform density is presumably the reason for the discrepancy between the MD simulation and Burgers's equation. 
At $A=10$ and $f=0.004$, the acoustic Mach number is $Ma=0.13$, which is outside the applicable region of Burgers's equation. 
An acoustic flow is generated at $Ma\simeq0.1$~\cite{iy94}.
Therefore, we conclude that higher-order terms of the continuum equation are required to reproduce the MD simulation results.

\begin{figure}[h]
  \includegraphics{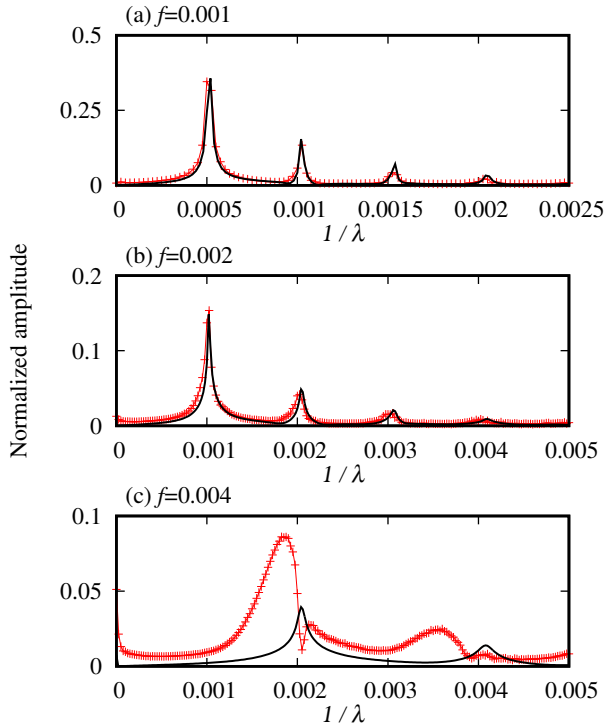}
  \caption{\label{fig:fft}
    Fourier transform of the waveform of the normalized velocity of the LJ fluid at $\rho=0.1$ and  $A=10$ for (a) $f=0.001$, (b) $f=0.002$, and (c) $f=0.004$. The red lines with ($+$) and the black lines represent the MD simulation results and the numerical solution of Burgers's equation, respectively.}
\end{figure}

Figure~\ref{fig:fft} shows the Fourier transform of the waveform, which can be used to determine the wavelength $\lambda$ of a soundwave. 
When the frequency $f$ and wavelength $\lambda$ are given, the speed of sound is $c = f\lambda$. 
The wavelength is obtained as the peak position of the fundamental wave $1/\lambda$.
The peak position is estimated by fitting the Gaussian function. 
The frequency of the sound source is adopted as the value of the frequency $f$.

At a low frequency ($f<0.002$), the results of the MD simulation and Burgers's equation agree very well, including the harmonic components. 
Hence, a quantitative agreement is obtained. 
However, at a high frequency ($f=0.004$), the position and height of the Fourier transform peaks are different. 
Therefore, the two waveforms are completely different.
This is because of the acoustic flow as described above.

\begin{figure}[h]
  \includegraphics{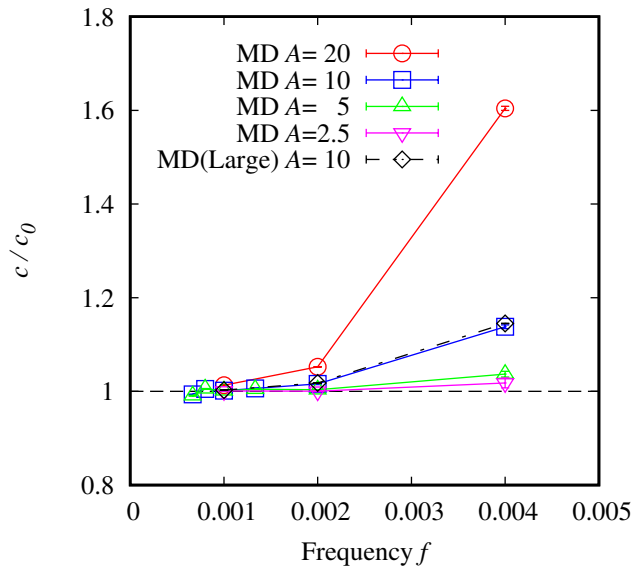}
  \caption{\label{fig:speed} Speed of sound in the LJ fluid at $\rho=0.1$.}
\end{figure}

Figure~\ref{fig:speed} shows the dependence of the speed of sound $c$ on the frequency at several amplitudes. 
In classical theory, the speed of sound is determined from the adiabatic bulk modulus, which is denoted as $c_0$.
At the low-frequency limit, the speed of sound converged to a value derived from classical theory independent of the amplitude.
As the frequency is increased, the speed of sound deviates from the value taken from classical theory. 
The speed of sound increases with the amplitude. The results of a larger system size with $L_y\times L_z = 25\times25$ are also included in Fig.~\ref{fig:speed}.
No significant differences are found due to different sizes.
Therefore, the size of the current simulation is sufficiently large to neglect the finite size effect.
The deviation cannot be explained solely by the dependence on $Ma$ because the data do not lie on a single curve when $c/c_0$ is plotted as a function of $Ma$.

\begin{figure}[h]
  \includegraphics{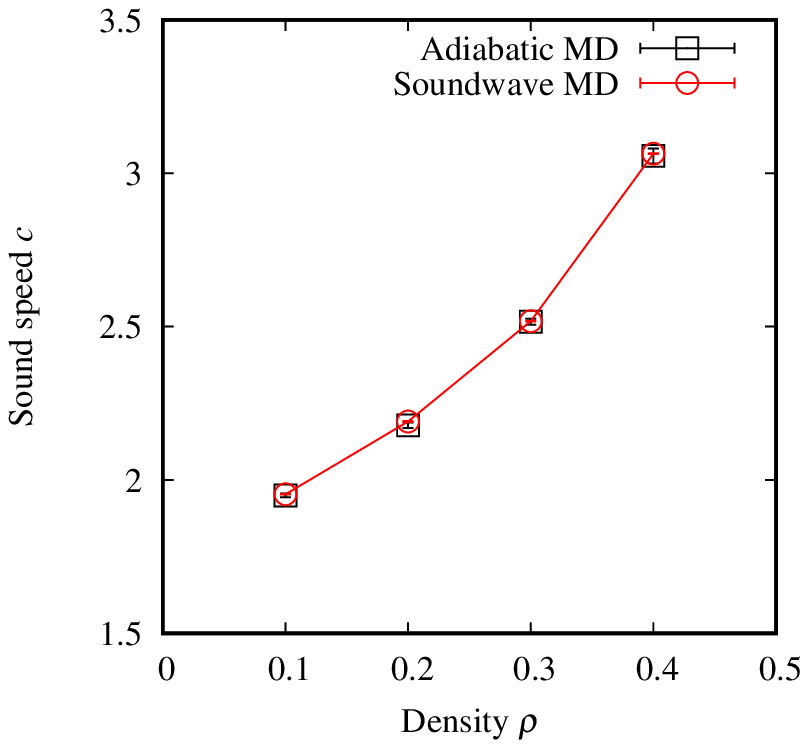}
  \caption{\label{fig:speed.T}
    Speed of sound in the LJ fluid as a function of density $\rho$.
    The circles represent the speeds calculated from the soundwave MD simulation
    with $A=2.5$ and $f=0.001$. The squares represent the speeds calculated from the adiabatic bulk modulus.}
\end{figure}

Figure~\ref{fig:speed.T} shows the dependence of the speed of sound on the density. 
The present method can also successfully measure the speed of sound in dense fluids.

\begin{figure}[h]
  \includegraphics{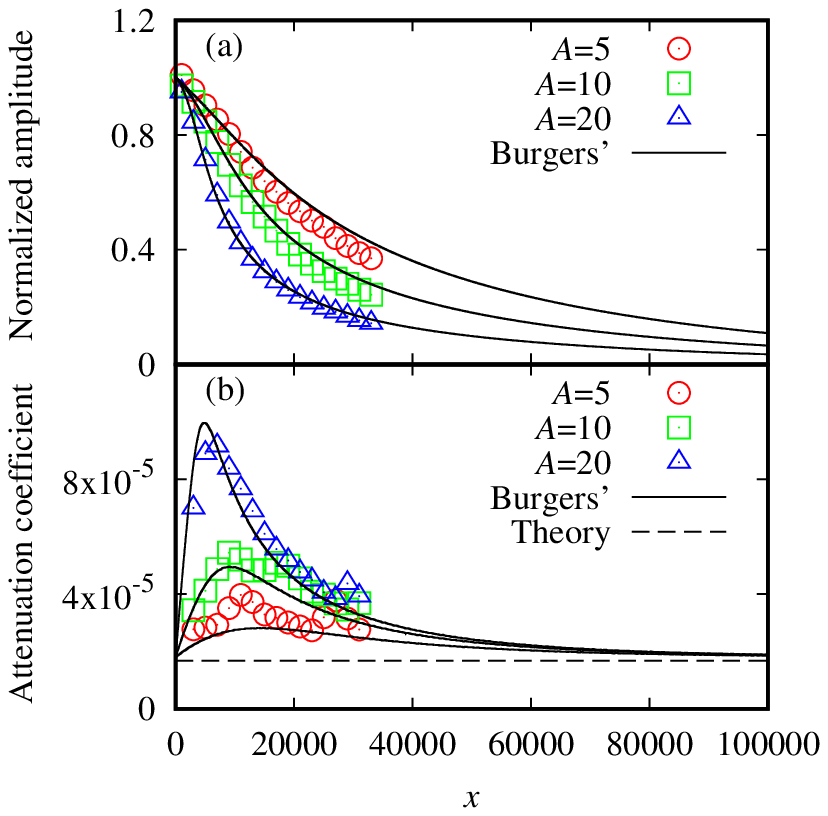}
  \caption{\label{fig:amplitude}
    (a) Normalized amplitude of the soundwave and (b) attenuation coefficient of the LJ fluid at $\rho=0.1$ as a function of $x$. The symbols and solid lines represent the results of the MD simulation and Burgers's equation, respectively, for the amplitudes $A=20$, $10$, and $5$ from top to bottom. The dashed line represents the attenuation coefficient from classical theory.
}
\end{figure}

The dissipation due to viscosity and heat conduction causes attenuation of the soundwave.
The attenuation behavior was also studied. 
At each position $x$, the amplitude is obtained through a Fourier transform of the variation in the flow velocity over time.
The amplitude is defined as the peak height of the fundamental wave obtained by the Fourier transform in the time direction.
Figure~\ref{fig:amplitude}(a) shows the spatial variation in the amplitude at $f=0.001$.
The MD simulation results indicate a moving average with a width of about a wavelength to reduce the influence of thermal fluctuations.
At small amplitudes, it is difficult to capture the attenuation characteristics away from the sound source because of thermal fluctuations.
Above a certain amplitude, the attenuation rates of the MD simulation and Burgers's equation are obtained consistently.

The attenuation coefficient $\alpha_0$ of the soundwaves is given by Eqs.~(\ref{eq:alpha}), (\ref{eq:c0}), and (\ref{eq:b}) at the low-frequency limit.
If the attenuation of the amplitude is assumed locally exponential, the attenuation coefficient $\alpha$ is given below,
\begin{eqnarray}
  \frac{{\rm d}a}{{\rm d}x}=-\alpha a,\label{eq:a}
\end{eqnarray}
where $a$ is the normalized amplitude.
Figure~\ref{fig:amplitude}(b) shows the spatial dependence of the attenuation coefficients obtained based on Eq.~(\ref{eq:a}).
The black lines show the results of the same calculations with Burgers's equation. 
The dashed line indicates the values from classical theory.
With Burgers's equation, the attenuation coefficient approaches the value of the classical theory sufficiently far from the sound source because the wave amplitude decreases. 
The coefficient also approaches the value of the classical theory at the oscillating plate ($x \simeq 0$).
This is likely because the waveform is close to the sinusoidal shape owing to the influence of the sinusoidal plate oscillation. 
With the MD simulation, the attenuation coefficient is difficult to estimate for the low amplitude  of $A=5$ because of thermal fluctuations.
In contrast, it quantitatively reproduces the results of Burgers's equation when the amplitude is $A>10$. 
Thus, the attenuation coefficient can be significantly overestimated if a large-amplitude wave is fitted, and it should be carefully calculated at the small-amplitude limit.

\section{\label{sec:SUMMARY AND DISCUSSION}SUMMARY AND DISCUSSION}
MD simulations were performed to analyze soundwaves generated from an oscillating plate in an LJ fluid.
Burgers's equation was used to derive a numerical solution for comparison. 
The results show that the waveforms are in a very good agreement at low frequencies. 
However, Burgers's equation is no longer applicable at high frequencies because of the acoustic flow near the oscillating plate.
The results demonstrated that MD simulations are applicable to a wider range of conditions.

The speed of sound is determined unambiguously, including the dependence on frequency by the Fourier transform of the waveform. 
However, it is difficult to obtain the attenuation coefficient from the MD simulation directly because of thermal fluctuations. 
There are two ways to address this problem. One is to increase the system size to remove the effect of thermal fluctuations.
However, performing an MD simulation at a size that is sufficiently large to confirm classical damping is impractical.
The other way is to make use of Burgers's equation. As shown in Fig.~\ref{fig:amplitude}, if the waveforms from the MD simulation and Burgers's equation match in a region close to the sound source, the latter should accurately estimate the behavior of the former away from the sound source.
Thus, the attenuation coefficient can be estimated by fitting a soundwave curve of Burgers's equation to that obtained by MD, when the coefficients in the Burgers's equation are unknown.

In this study, all parameters required for Burgers's equation were obtained from MD simulations, so no fitting parameters were used.
However, determining the parameters of Burgers's or another continuum equation from MD simulations is difficult for general complex fluids.
Therefore, the parameters of a continuum equation can be determined by fitting the shapes of the soundwaves from the MD simulation.
Because obtaining the physical quantities necessary for calculating the damping rate is difficult with general complex fluids, especially the volumetric viscosity, this method is useful for estimating the damping rate of complex systems.

We conclude that MD simulations are useful for the quantitative analysis of soundwaves. They allow the fluid dynamics to be discussed without requiring assumptions for the local equilibrium. Moreover, the relation between the molecular structure and fluid properties can be investigated in detail. They are a promising tool for the analysis of complex fluids, such as gas-liquid two-phase flow involving phase transitions, where a continuum description is not appropriate.

\begin{acknowledgments}
  We thank Y. Higuchi for helpful discussions. 
  This research was supported by MEXT as ``Exploratory Challenge on Post-K computer'' (Challenge of Basic Science¡½Exploring Extremes through Multi-Physics and Multi-Scale Simulations) and JSPS KAKENHI (Grant No. JP19H05718).
  Computation was partially carried out by using the facilities of the Supercomputer Center, Institute for Solid State Physics, University of Tokyo, and Research Center for Computational Science, Okazaki, Japan.
\end{acknowledgments}

\section*{DATA AVAILABILITY}
The data that support the findings of this study are available from the corresponding author upon reasonable request.
\appendix

\section{Parameters in the Burgers's equation}

Table~\ref{table:prm} summarizes the parameter values included in Burgers's equation for the LJ fluid (see Eq.~(\ref{eq:burgers})). Each parameter was estimated through MD simulations, and the calculation methods are described in below.
\begin{table*}[t]
  \caption{Parameter values included in Burgers's equation for the LJ fluid. The number in the brackets is the accuracy of the last digit.}
  \label{table:prm}
  \centering
  \begin{tabular}{cccccccc}
    \hline
    $c_0$ & $c_{V}$ & $c_{p}$ & $\eta$ & $\zeta$ & $\kappa$ & $B_1$   & $B_2$   \\
    \hline
    1.950(5)  & 1.580(7)    & 3.07(3)    & 0.244(2)  & 0.013(1)  & 1.104(4)     & 0.3811(3) & 0.45(4) \\
    \hline
  \end{tabular}
\end{table*}

\subsection{Speed of sound and specific heat}
The speed of sound $c_0$ was estimated from Eq.~(\ref{eq:c0}).
To determine the adiabatic bulk modulus, we adopted the following thermodynamic relationship:
\begin{eqnarray}
  K_S=\frac{c_p}{c_V}K_T,
\end{eqnarray}
where $c_p$, $c_V$, and $K_T$ are the isobaric specific heat, isochoric specific heat, and isothermal bulk modulus, respectively.
These three quantities are obtained from the fluctuations of physical quantities,
\begin{eqnarray}
  c_p &=& \frac{1}{T^2}\left( \left\langle h^2\right\rangle_{NpT} - \left\langle h \right\rangle_{NpT}^2 \right),\\
  c_V &=& \frac{1}{T^2}\left( \left\langle e^2\right\rangle_{NVT} - \left\langle e \right\rangle_{NVT}^2 \right), \\
  \frac{1}{K_T} &=& \frac{1}{T\left\langle V\right\rangle_{NpT}}\left( \left\langle V^2 \right\rangle_{NpT} - \left\langle V \right\rangle_{NpT}^2 \right),
\end{eqnarray}
where $h$, $e$, and $V$ are the enthalpy per atom,energy per atom, and volume of the system, respectively.
$\langle\cdot\rangle_{NpT}$ and $\langle\cdot\rangle_{NVT}$ represent the statistical averages with the isothermal-isobaric ensemble and isothermal-isochoric ensemble, respectively. 
These statistical averages were estimated through MD simulations. 

The Nose--Hoover barostat~\cite{hoover86} and Langevin thermostat are used for pressure control and temperature control, respectively.
The simulation box is a cube with an edge length of $L=100$.

\subsection{Shear viscosity}
\begin{figure}[h]
  \includegraphics{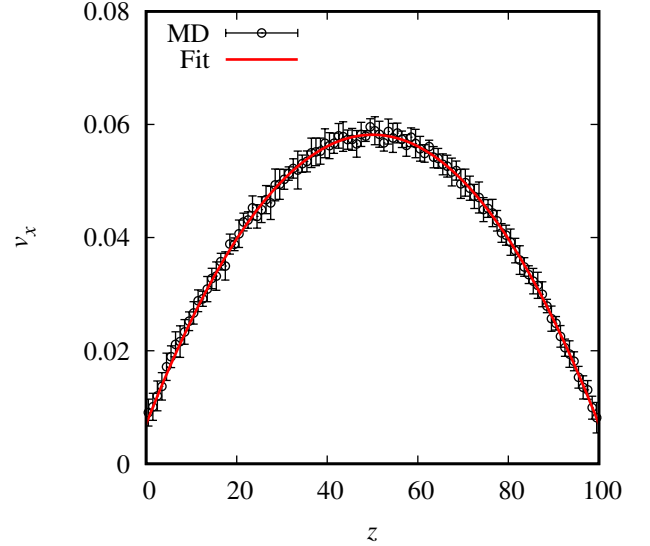}
  \caption{\label{fig:a1} Fitting results with Eq.~(\ref{eq:eta}) for the Poiseuille flow of the LJ fluid at $\rho = 0.1$.}
\end{figure}

The shear viscosity $\eta$ of the LJ fluid was estimated by generating a Poiseuille flow in the MD simulation~\cite{awn20}.
The flow is generated by imposing a gravitational acceleration $g=0.0001$ in the $x$-direction.
The simulation box is a rectangular parallelepiped with the dimensions of $L_x\times L_y\times L_z=100\times 100\times 120$.
The wall is modeled by using a Langevin thermostat in the region of $100<z<120$.
The shear viscosity coefficient $\eta$ is obtained by fitting the $x$-component of the flow velocity $v_x(z)$ to the following equation:
\begin{eqnarray}
  v_x(z)=\frac{\rho g}{2\eta}z(100-z)+v_{x0},\label{eq:eta}
\end{eqnarray}
where $v_{x0}$ is the slip velocity at the boundary of the thermostat region. Here, $v_{x0}$ is treated as a fitting parameter.
Figure~\ref{fig:a1} shows the fitting results.

\subsection{Bulk viscosity}
\begin{figure}[h]
  \includegraphics{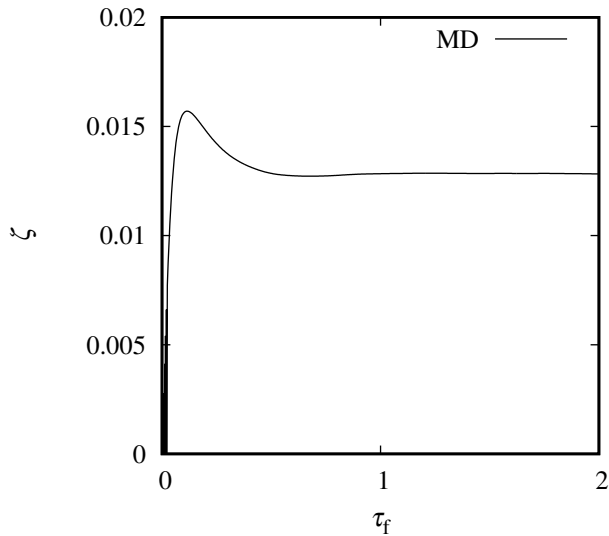}
  \caption{\label{fig:a2} Bulk viscosity $\zeta$ of the LJ fluid at $\rho=0.1$. 
The upper limit of the integral in Eq.~(\ref{eq:zeta}) is replaced with $\tau_{\rm f}$.}
\end{figure}

The bulk viscosity $\zeta$ of the LJ fluid was estimated with the Green-Kubo formula~\cite{zwanzig65}:
\begin{eqnarray}
  \zeta &=& \frac{V}{T}\int_{0}^{\infty}{\rm d}t \left\langle \left(p(t)-\langle p\rangle \right) \left(p(0) - \langle p\rangle \right)  \right\rangle,\label{eq:zeta}
\end{eqnarray}
where $\langle\cdots\rangle$ denotes the statistical average in the microcanonical ensemble.
The statistical average was estimated through an MD simulation. 
The simulation box is a cube with an edge length of $L=100$.
Figure~\ref{fig:a2} shows the behavior of $\zeta$ as the upper limit  $\tau_{\rm f}$ of the integral in Eq.~(\ref{eq:zeta}) is increased.
A cutoff is made where the value of the integral reached a constant value.
\begin{figure}[h]
  \includegraphics{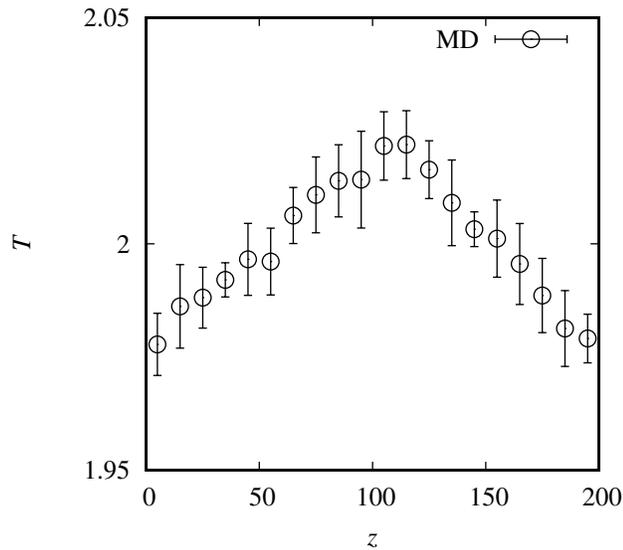}
  \caption{\label{fig:a3} Temperature profile generated by the particle exchange method at $\rho=0.1$.}
\end{figure}

\subsection{Thermal conductivity}
The thermal conductivity $\kappa$ was evaluated by generating a temperature gradient and heat flow with the particle exchange method~\cite{muller97} in an MD simulation.
The simulation box is a rectangular parallelepiped with the dimensions of $L_x\times L_y\times L_z=100\times 100 \times 200$.
The hot region is defined as $100<z<110$, and the cold region is defined as $0<z<10$. 
The velocities of particles with the lowest kinetic energy in the hot region and the highest kinetic energy in the cold region are exchanged every $500$ steps.
The energy transfer $\Delta Q$ from the velocity exchange is balanced by the heat flow from the temperature gradient $\Delta T/\Delta z$, which generates a steady state.
Figure~\ref{fig:a3} shows the temperature profile generated by the particle exchange method.
The thermal conductivity $\kappa$ is given as follows:
\begin{eqnarray}
  \kappa &=& -\frac{\Delta Q}{2t_{\rm sim}L_xL_y\frac{\Delta T}{\Delta z}},
\end{eqnarray}
where $t_{\rm sim}$ is the simulation time.

\subsection{Nonlinear parameter}
Equations~(\ref{eq:A}) and (\ref{eq:B}) define the first and second adiabatic differential coefficients of pressures $B_1$ and $B_2$. These coefficients were estimated according to the adiabatic changes in the MD simulation. First, pressure $p$ is calculated at $T=2$ and $\rho=0.1$.
For the numerical differentiation, the pressures at the densities $\rho+\Delta \rho$ and $\rho-\Delta \rho$ are calculated, where $\Delta \rho=0.01$ is used. The temperature is adjusted so that the total energy change of the system satisfies the relation $\Delta e = p\Delta \rho/\rho^2$.
The central difference method is used to determine  $B_1$ and $B_2$ from the obtained pressure.
The simulation box  is a cube with an edge length of $L=100$.
\nocite{*}

\providecommand{\noopsort}[1]{}\providecommand{\singleletter}[1]{#1}%

\end{document}